\documentstyle[10pt,epsf,epsfig,dp_delphititle,here]{dp_delphi}
\makeindex
\def\DpTitle{{
Determination of the $e^+e^- \to \gamma \gamma (\gamma)$ 
cross-section  
at centre-of-mass energies ranging from \\ 189 GeV to 202 GeV }}
\def\DpPaperGroup{EP}
\def\DpPaperRef{2000-094}
\def\DpDate{13 July 2000}
\def\DpAuthors{DELPHI Collaboration}
\def\DpSubmit{(Submitted to Physics Letters B)}
\def\DpComment{ }
\def\DpEMail{ }

\newcommand{\eeto} {e^+e^- \to}

\newcommand{\gr} {\mbox{$^{\circ}$}}

\newcommand{\fot} {\mbox{$\gamma$}}

\newcommand{\eeintogg} {\mbox{$\eeto \, \gamma \gamma$}}

\newcommand{\auth}[1]{#1,}

\newcommand{\etal}{et al.}

\newcommand{\J}[4]{{#1}{\bf #2}, #3 (#4)}

\newcommand{\PLB}[3]{\J{Phys.\,Lett.\,B}{#1}{#2}{#3}}

\newcommand{\npB}[3]{{\sl Nucl. Phys.} {\bf B#1} (19#2)~#3}

\newcommand{\prl}[3]{{\sl Phys. Rev. Lett.} {\bf } (19#2) #3}

\newcommand{\EPJ}[3]{{\sl E.Phys.J.} {\bf C#1} (19#2) #3}
\newcommand{\nim}[3]{{\sl Nucl.\,Instrum.\,Methods} {\bf A#1} (19#2) #3}
\begin{document}
\makeatletter
\newcount\@tempcntc
\def\@citex[#1]#2{\if@filesw\immediate\write\@auxout{\string\citation{#2}}\fi
  \@tempcnta\z@\@tempcntb\m@ne\def\@citea{}\@cite{\@for\@citeb:=#2\do
    {\@ifundefined
       {b@\@citeb}{\@citeo\@tempcntb\m@ne\@citea\def\@citea{,}{\bf ?}\@warning
       {Citation `\@citeb' on page \thepage \space undefined}}%
    {\setbox\z@\hbox{\global\@tempcntc0\csname b@\@citeb\endcsname\relax}%
     \ifnum\@tempcntc=\z@ \@citeo\@tempcntb\m@ne
       \@citea\def\@citea{,}\hbox{\csname b@\@citeb\endcsname}%
     \else
      \advance\@tempcntb\@ne
      \ifnum\@tempcntb=\@tempcntc
      \else\advance\@tempcntb\m@ne\@citeo
      \@tempcnta\@tempcntc\@tempcntb\@tempcntc\fi\fi}}\@citeo}{#1}}
\def\@citeo{\ifnum\@tempcnta>\@tempcntb\else\@citea\def\@citea{,}%
  \ifnum\@tempcnta=\@tempcntb\the\@tempcnta\else
   {\advance\@tempcnta\@ne\ifnum\@tempcnta=\@tempcntb \else \def\@citea{--}\fi
    \advance\@tempcnta\m@ne\the\@tempcnta\@citea\the\@tempcntb}\fi\fi}
 
\makeatother
\begin{titlepage}
\pagenumbering{roman}
\CERNpreprint{\DpPaperGroup}{\DpPaperRef} 
\date{{\small\DpDate}} 
\title{\DpTitle} 
\address{\DpAuthors} 
\begin{shortabs} 
\noindent
%
A test of the QED process \eeintogg(\fot) is reported.
The data analysed were collected with the DELPHI detector
in 1998 and 1999 at the highest ener\-gies achieved at LEP, 
reaching 202 GeV in the centre-of-mass.
The total integrated luminosity amounts to 375.7 pb$^{-1}$.
The differential and total cross-sections
for the process $e^+e^- \to \gamma \gamma$ were measured,
and found to be in agreement with the QED prediction. 95\% 
Confidence Level (C.L.) lower limits on the QED cut-off
parameters of $\rm \Lambda_+ >$ 330 GeV and  $\rm \Lambda_{-}>$ 320 GeV 
were derived.  A 95\% 
C.L. lower bound on the mass of an excited electron of 
311 $\rm GeV/c^2$ (for $\lambda_{\gamma}$ =1) was obtained. 
s-channel virtual graviton exchange was searched for, resulting in 95\% 
C.L. lower limits on the string mass scale,
$\rm M_S$:  
$\rm M_S > 713 \, GeV/c^2$ ( $\lambda=1$) and 
$\rm M_S > 691 \,GeV/c^2$ ($\lambda = -1$). 

\end{shortabs}
\vfill
\begin{center}
\DpSubmit \ \\ 
\DpComment \ \\
\DpEMail \ \\
\end{center}
\vfill
\clearpage
\begingroup
%
\newcommand{\DpName}[2]{\hbox{#1$^{\ref{#2}}$},\hfill}
\newcommand{\DpNameTwo}[3]{\hbox{#1$^{\ref{#2},\ref{#3}}$},\hfill}
\newcommand{\DpNameThree}[4]{\hbox{#1$^{\ref{#2},\ref{#3},\ref{#4}}$},\hfill}
\newskip\Bigfill \Bigfill = 0pt plus 1000fill
\newcommand{\DpNameLast}[2]{\hbox{#1$^{\ref{#2}}$}\hspace{\Bigfill}}
\footnotesize
\noindent
\DpName{P.Abreu}{LIP}
\DpName{W.Adam}{VIENNA}
\DpName{T.Adye}{RAL}
\DpName{P.Adzic}{DEMOKRITOS}
\DpName{Z.Albrecht}{KARLSRUHE}
\DpName{T.Alderweireld}{AIM}
\DpName{G.D.Alekseev}{JINR}
\DpName{R.Alemany}{VALENCIA}
\DpName{T.Allmendinger}{KARLSRUHE}
\DpName{P.P.Allport}{LIVERPOOL}
% ...
\normalsize
\titlefoot{Department of Physics and Astronomy, Iowa State
     University, Ames IA 50011-3160, USA
    \label{AMES}}
\titlefoot{Physics Department, Univ. Instelling Antwerpen,
     Universiteitsplein 1, B-2610 Antwerpen, Belgium \\
     \indent~~and IIHE, ULB-VUB,
     Pleinlaan 2, B-1050 Brussels, Belgium \\
     \indent~~and Facult\'e des Sciences,
     Univ. de l'Etat Mons, Av. Maistriau 19, B-7000 Mons, Belgium
    \label{AIM}}
\titlefoot{Physics Laboratory, University of Athens, Solonos Str.
     104, GR-10680 Athens, Greece
    \label{ATHENS}}
\titlefoot{Department of Physics, University of Bergen,
     All\'egaten 55, NO-5007 Bergen, Norway
    \label{BERGEN}}
\titlefoot{Dipartimento di Fisica, Universit\`a di Bologna and INFN,
     Via Irnerio 46, IT-40126 Bologna, Italy
    \label{BOLOGNA}}
\titlefoot{Centro Brasileiro de Pesquisas F\'{\i}sicas, rua Xavier Sigaud 150,
     BR-22290 Rio de Janeiro, Brazil \\
     \indent~~and Depto. de F\'{\i}sica, Pont. Univ. Cat\'olica,
     C.P. 38071 BR-22453 Rio de Janeiro, Brazil \\
     \indent~~and Inst. de F\'{\i}sica, Univ. Estadual do Rio de Janeiro,
     rua S\~{a}o Francisco Xavier 524, Rio de Janeiro, Brazil
    \label{BRASIL}}
\titlefoot{Comenius University, Faculty of Mathematics and Physics,
     Mlynska Dolina, SK-84215 Bratislava, Slovakia
    \label{BRATISLAVA}}
\titlefoot{Coll\`ege de France, Lab. de Physique Corpusculaire, IN2P3-CNRS,
     FR-75231 Paris Cedex 05, France
    \label{CDF}}
\titlefoot{CERN, CH-1211 Geneva 23, Switzerland
    \label{CERN}}
\titlefoot{Institut de Recherches Subatomiques, IN2P3 - CNRS/ULP - BP20,
     FR-67037 Strasbourg Cedex, France
    \label{CRN}}
\titlefoot{Now at DESY-Zeuthen, Platanenallee 6, D-15735 Zeuthen, Germany
    \label{DESY}}
\titlefoot{Institute of Nuclear Physics, N.C.S.R. Demokritos,
     P.O. Box 60228, GR-15310 Athens, Greece
    \label{DEMOKRITOS}}
\titlefoot{FZU, Inst. of Phys. of the C.A.S. High Energy Physics Division,
     Na Slovance 2, CZ-180 40, Praha 8, Czech Republic
    \label{FZU}}
\titlefoot{Dipartimento di Fisica, Universit\`a di Genova and INFN,
     Via Dodecaneso 33, IT-16146 Genova, Italy
    \label{GENOVA}}
\titlefoot{Institut des Sciences Nucl\'eaires, IN2P3-CNRS, Universit\'e
     de Grenoble 1, FR-38026 Grenoble Cedex, France
    \label{GRENOBLE}}
\titlefoot{Helsinki Institute of Physics, HIP,
     P.O. Box 9, FI-00014 Helsinki, Finland
    \label{HELSINKI}}
\titlefoot{Joint Institute for Nuclear Research, Dubna, Head Post
     Office, P.O. Box 79, RU-101 000 Moscow, Russian Federation
    \label{JINR}}
\titlefoot{Institut f\"ur Experimentelle Kernphysik,
     Universit\"at Karlsruhe, Postfach 6980, DE-76128 Karlsruhe,
     Germany
    \label{KARLSRUHE}}
\titlefoot{Institute of Nuclear Physics and University of Mining and Metalurgy,
     Ul. Kawiory 26a, PL-30055 Krakow, Poland
    \label{KRAKOW}}
\titlefoot{Universit\'e de Paris-Sud, Lab. de l'Acc\'el\'erateur
     Lin\'eaire, IN2P3-CNRS, B\^{a}t. 200, FR-91405 Orsay Cedex, France
    \label{LAL}}
\titlefoot{School of Physics and Chemistry, University of Lancaster,
     Lancaster LA1 4YB, UK
    \label{LANCASTER}}
\titlefoot{LIP, IST, FCUL - Av. Elias Garcia, 14-$1^{o}$,
     PT-1000 Lisboa Codex, Portugal
    \label{LIP}}
\titlefoot{Department of Physics, University of Liverpool, P.O.
     Box 147, Liverpool L69 3BX, UK
    \label{LIVERPOOL}}
\titlefoot{LPNHE, IN2P3-CNRS, Univ.~Paris VI et VII, Tour 33 (RdC),
     4 place Jussieu, FR-75252 Paris Cedex 05, France
    \label{LPNHE}}
\titlefoot{Department of Physics, University of Lund,
     S\"olvegatan 14, SE-223 63 Lund, Sweden
    \label{LUND}}
\titlefoot{Universit\'e Claude Bernard de Lyon, IPNL, IN2P3-CNRS,
     FR-69622 Villeurbanne Cedex, France
    \label{LYON}}
\titlefoot{Univ. d'Aix - Marseille II - CPP, IN2P3-CNRS,
     FR-13288 Marseille Cedex 09, France
    \label{MARSEILLE}}
\titlefoot{Dipartimento di Fisica, Universit\`a di Milano and INFN-MILANO,
     Via Celoria 16, IT-20133 Milan, Italy
    \label{MILANO}}
\titlefoot{Dipartimento di Fisica, Univ. di Milano-Bicocca and
     INFN-MILANO, Piazza delle Scienze 2, IT-20126 Milan, Italy
    \label{MILANO2}}
\titlefoot{Niels Bohr Institute, Blegdamsvej 17,
     DK-2100 Copenhagen {\O}, Denmark
    \label{NBI}}
\titlefoot{IPNP of MFF, Charles Univ., Areal MFF,
     V Holesovickach 2, CZ-180 00, Praha 8, Czech Republic
    \label{NC}}
\titlefoot{NIKHEF, Postbus 41882, NL-1009 DB
     Amsterdam, The Netherlands
    \label{NIKHEF}}
\titlefoot{National Technical University, Physics Department,
     Zografou Campus, GR-15773 Athens, Greece
    \label{NTU-ATHENS}}
\titlefoot{Physics Department, University of Oslo, Blindern,
     NO-1000 Oslo 3, Norway
    \label{OSLO}}
\titlefoot{Dpto. Fisica, Univ. Oviedo, Avda. Calvo Sotelo
     s/n, ES-33007 Oviedo, Spain
    \label{OVIEDO}}
\titlefoot{Department of Physics, University of Oxford,
     Keble Road, Oxford OX1 3RH, UK
    \label{OXFORD}}
\titlefoot{Dipartimento di Fisica, Universit\`a di Padova and
     INFN, Via Marzolo 8, IT-35131 Padua, Italy
    \label{PADOVA}}
\titlefoot{Rutherford Appleton Laboratory, Chilton, Didcot
     OX11 OQX, UK
    \label{RAL}}
\titlefoot{Dipartimento di Fisica, Universit\`a di Roma II and
     INFN, Tor Vergata, IT-00173 Rome, Italy
    \label{ROMA2}}
\titlefoot{Dipartimento di Fisica, Universit\`a di Roma III and
     INFN, Via della Vasca Navale 84, IT-00146 Rome, Italy
    \label{ROMA3}}
\titlefoot{DAPNIA/Service de Physique des Particules,
     CEA-Saclay, FR-91191 Gif-sur-Yvette Cedex, France
    \label{SACLAY}}
\titlefoot{Instituto de Fisica de Cantabria (CSIC-UC), Avda.
     los Castros s/n, ES-39006 Santander, Spain
    \label{SANTANDER}}
\titlefoot{Dipartimento di Fisica, Universit\`a degli Studi di Roma
     La Sapienza, Piazzale Aldo Moro 2, IT-00185 Rome, Italy
    \label{SAPIENZA}}
\titlefoot{Inst. for High Energy Physics, Serpukov
     P.O. Box 35, Protvino, (Moscow Region), Russian Federation
    \label{SERPUKHOV}}
\titlefoot{J. Stefan Institute, Jamova 39, SI-1000 Ljubljana, Slovenia
     and Laboratory for Astroparticle Physics,\\
     \indent~~Nova Gorica Polytechnic, Kostanjeviska 16a, SI-5000 Nova Gorica, Slovenia, \\
     \indent~~and Department of Physics, University of Ljubljana,
     SI-1000 Ljubljana, Slovenia
    \label{SLOVENIJA}}
\titlefoot{Fysikum, Stockholm University,
     Box 6730, SE-113 85 Stockholm, Sweden
    \label{STOCKHOLM}}
\titlefoot{Dipartimento di Fisica Sperimentale, Universit\`a di
     Torino and INFN, Via P. Giuria 1, IT-10125 Turin, Italy
    \label{TORINO}}
\titlefoot{Dipartimento di Fisica, Universit\`a di Trieste and
     INFN, Via A. Valerio 2, IT-34127 Trieste, Italy \\
     \indent~~and Istituto di Fisica, Universit\`a di Udine,
     IT-33100 Udine, Italy
    \label{TU}}
\titlefoot{Univ. Federal do Rio de Janeiro, C.P. 68528
     Cidade Univ., Ilha do Fund\~ao
     BR-21945-970 Rio de Janeiro, Brazil
    \label{UFRJ}}
\titlefoot{Department of Radiation Sciences, University of
     Uppsala, P.O. Box 535, SE-751 21 Uppsala, Sweden
    \label{UPPSALA}}
\titlefoot{IFIC, Valencia-CSIC, and D.F.A.M.N., U. de Valencia,
     Avda. Dr. Moliner 50, ES-46100 Burjassot (Valencia), Spain
    \label{VALENCIA}}
\titlefoot{Institut f\"ur Hochenergiephysik, \"Osterr. Akad.
     d. Wissensch., Nikolsdorfergasse 18, AT-1050 Vienna, Austria
    \label{VIENNA}}
\titlefoot{Inst. Nuclear Studies and University of Warsaw, Ul.
     Hoza 69, PL-00681 Warsaw, Poland
    \label{WARSZAWA}}
\titlefoot{Fachbereich Physik, University of Wuppertal, Postfach
     100 127, DE-42097 Wuppertal, Germany
    \label{WUPPERTAL}}
\endgroup
\end{titlepage}
\clearpage
%
\pagenumbering{arabic} 
\setcounter{footnote}{0} %
\large

\section{Introduction}

An analysis of two-photon final states using the high energy 
data sets collected
with the DELPHI detector in 1998 and 1999 is reported.
The data analysed were collected at $e^+ e^-$ 
collision energies ranging from 188.6 GeV up to 201.6 GeV, 
corresponding to a total integrated luminosity of 375.7 pb$^{-1}$.

Final states with two photons are mainly produced by 
the standard process $e^+e^- \to \gamma \gamma (\gamma)$.
This reaction is an almost pure QED process:
at orders above $\alpha^2$, it is mainly affected 
by QED corrections, such as soft and hard {\it bremsstrahlung}
and virtual corrections, compared to which the  
weak corrections due to the exchange
of virtual massive gauge bosons are very small 
\cite{bib_BK,bib_bohm,bib_fuji}.
Therefore, any significant deviation between the measured and the QED 
cross-section could unambiguously be interpreted as the result of 
non-standard physics.

The Born cross-section for $e^+e^- \to \gamma \gamma (\gamma)$ is given by

\begin{equation}
\rm \sigma^{0}_{QED}=K \cdot \frac{2 \pi \alpha}{s}.
\label{eq_sqed}
\end{equation}

\noindent
K depends on the angular acceptance for the final state photons,
$\rm \alpha$ is the electromagnetic coupling constant and s is the 
centre-of-mass energy squared. 

Since $\rm \sigma^{0}_{QED}$ scales with $\rm s^{-1}$, the combination of 
measurements taken at different centre-of-mass energy values is 
straightforward and data taken at neighbouring
values of $\sqrt{s}$ can be combined by applying this scaling function.  

Previous DELPHI results concerning the process
$e^+e^- \to \gamma \gamma (\gamma)$, using LEPI and LEPII data, 
can be found in references \cite{bib_94,bib_97}. 
The most recently published results from the other LEP experiments
can be found in references \cite{bib_aleph,bib_l3,bib_opal}.

\section{Data sample and apparatus}

The data analysed were taken at $e^+ e^-$ collision energies
of 188.63 $\pm$ 0.04 GeV, 191.6 $\pm$ 0.04 GeV, 195.5 $\pm$ 0.04 GeV, 
199.5 $\pm$ 0.04 GeV and 201.6 $\pm$ 0.04 GeV \cite{bib_energy},
corresponding to integrated luminosities of 151.9 $\pm$ 0.9 pb$^{-1}$, 
25.1 $\pm$ 0.1 pb$^{-1}$, 76.1 $\pm$ 0.4 pb$^{-1}$, 82.6 $\pm$ 0.5 pb$^{-1}$ and 40.1 $\pm$ 0.2 pb$^{-1}$
respectively.
The luminosity was measured by counting the number of Bhabha events
at small polar angles, recorded with DELPHI's luminometer: 
the Small angle TIle Calorimeter (STIC),
made of two modules located at $|z|$ = 220 cm 
from the interaction point and with polar angle
coverage between 2\gr\ and 10\gr\ (170\gr\ and 178\gr ).

Photon detection and reconstruction relies on the trigger 
and energy measurement based on two electro\-magnetic 
calorimeters: the High density Projection Chamber (HPC) 
in the barrel region and 
the Forward ElectroMagnetic Calorimeter (FEMC) in the endcaps.
The HPC is a gas-sampling calorimeter, made of 144 modules,
each one with 10 lead layers in $\rm R\phi$ embedded in a gas mixture.
It covers polar angles between 42\gr\ and 138\gr . 
The FEMC is a lead glass calorimeter, 
covering the polar angle region $[11^\circ,35^\circ]$
and its complement with respect to 180\gr. 
The barrel DELPHI electro\-magne\-tic trigger 
requires coincidence between scintillator signals 
and energy deposits in HPC while in the forward 
region the electromagnetic trigger is given 
by energy deposits in the FEMC lead-glass counters.

The tracking system allows the rejection of charged particles and the
recovery of photons converting inside the detector.
The DELPHI barrel tracking
system relies on the Vertex Detector (VD), the Inner Detector (ID),
the Time Projection Chamber (TPC) and the Outer Detector (OD).
In the endcaps, the tracking system relies also on the VD and the TPC
(down to about 20\gr\ in polar angle),
and on the Forward Chambers A and B (FCA, FCB).
The VD plays an important role in the
detection of charged particle tracks coming from the interaction
point.
A more detailed description of the DELPHI detector, of the 
triggering conditions and of the readout chain can be found 
in \cite{bib_delphi}. 

\section{Photon reconstruction and identification}

The process \eeintogg(\fot) yields not only neutral final states 
but also final states cha\-racte\-rized by 
the presence of charged particle tracks from photon conversions.

Photons converting inside the tracking 
system, but after the Vertex Detector, are characterized by charged 
particle tracks and will be refered to as converted photons.
Photons reaching the electromagnetic calorimeters
before converting, yielding no reconstructed charged particle tracks,
will be refered to as unconverted photons.
According to this classification, 
two different algorithms were 
applied in the photon reconstruction and identification.


  
The main contamination to \eeintogg(\fot) final states comes from 
radiative Bhabha ($e^+e^- \to e^+e^-(\gamma)$) 
events with one non-reconstructed electron and the other electron 
lost in the beam pipe, and from Compton ($e^\pm \gamma$) events. 
Compton events are produced by the scattering of beam electrons
by a quasi-real photon radiated by another incoming electron,   
resulting mostly in final states with one photon and one electron 
in the detector, the remaining $e^\pm$ going undetected through the beam-pipe.
Both the Bhabha and the Compton backgrounds 
can however be dramatically reduced if the Vertex Detector is used as a veto 
for charged particles coming from the interaction point. 
The event generator used to simulate \eeintogg(\fot) was that 
of Berends and Kleiss \cite{bib_BK}, 
while the Bhabha and Compton event generators are BHWIDE and TEEG, 
described in references \cite{bib_bhwide} and \cite{bib_compton} respectively.
The generated samples were  processed through the full DELPHI 
simulation and reconstruction chains \cite{bib_delphi}.

\subsection{Unconverted photons \label{sec:unconverted}}

Unconverted photon candidates were reconstructed by applying 
an isolation algorithm to energy deposits in the calorimeters.
The algorithm relied on a double cone centered on
each energy deposit, with internal and external half angles of 5\gr\ 
and 15\gr\ respectively, where the vertices of 
both cones correspond to the geometric centre of DELPHI.
Showers were considered isolated if the total energy inside the double 
cone was less than 1 GeV.
The energy of the isolated neutral particles was re-evaluated as 
the sum of the energy of all associated deposits inside the inner cone
where no charged particles of more than 250 MeV/c were allowed.
The direction of the isolated showers was the
energy weighted mean of the directions of all associated 
energy deposits.
Such particles, with a total energy above 3 GeV, were 
identified as photons if the following criteria were fulfilled:

\begin{itemize}
\item 
The polar angle of the energy deposit was inside  
$[25^{\circ},35^{\circ}]$, 
$[42^{\circ},88^{\circ}]$,  
$ [92^{\circ},138^{\circ}]$ or 
$ [145^{\circ},155^{\circ}]$, in order to reduce VD and calorimeter
edge effects.

\item 
No VD track element 
pointed to the direction of the energy deposit 
within 3$^{\circ}$ (10$^{\circ}$) in azimuthal
angle in the barrel (forward) region of DELPHI
(a VD track element was defined as
at least two hits in different VD layers
aligned within an azimuthal angle interval of 0.5$^{\circ}$).

\item 
If more than 3 GeV of hadronic energy was 
associated to a deposit, then at least 90\% 
of it had to be in the first layer of the 
Hadronic CALorimeter (HCAL).

\item 
For an energy deposit in the HPC,
there had to be at least three HPC layers with more than 5\% 
of the total electromagnetic energy, unless  
the deposit was within 1 degree of the 
HPC azimuthal intermodular divisions 
\footnote{The HPC modules are distributed in 6 rings of 
24 modules located at mod$(\phi,15^{\circ}) = 7.5^\circ$.}.

\end{itemize}

\subsection{Converted photons\label{sec_cp}}

Converted photon candidates were reconstructed with the help of 
a jet clustering algorithm: all particles in the event,
with the exception of isolated neutral particles,
were forced to be clustered in jets
(isolated charged particles were not treated as 
single particles but as low multiplicity jets).
The DURHAM jet algorithm \cite{bib_durham} was applied, using as resolution 
variable $y_{cut}=0.003$.
Low multiplicity jets with less than 6 charged particles 
were treated as converted photon candidates.
These candidates were recovered 
if they were associated to energy deposits above 3 GeV
fulfilling the photon identification criteria 
described in section \ref{sec:unconverted}.
The requirement that no correlated signals were observed in the VD 
was a particularly important criterion for the rejection of electrons.

\section{Two photon events: {\bf \eeintogg (\fot) }}

The selected \fot \fot(\fot) sample consisted of events
with at least two photons, where at most one was converted.
The electromagnetic calorimeters (HPC and FEMC), the 
TPC and the VD were required to be nominally operational. 
The analysis was performed in the polar angle interval 
corresponding to 
${\rm |\cos{\theta}^{\ast}| \in [0.035,0.731] \cup [0.819,0.906]}$,
where the variable $\theta^{\ast}$ stands for  
the polar angle of the photons 
relative to the direction of the incident electron
in the centre-of-mass of the $e^+e^-$ collision\footnote
{The parameterization of the photon polar angle
with ${\rm \theta^{\ast}}$ enables 
the cross-section measurement to be insensitive to photons 
lost in the beam pipe.} after allowing for ISR.
The two most energetic photons were required to have energies above 15\% 
of the collision energy and isolation angle of 30$^{\circ}$
(the isolation angle is the minimum of the angles between the 
photon and the remaining reconstructed particles in the event).
No other particles (with exception of isolated photons) 
with energy above 3 GeV were allowed in the event.
The application of these criteria resulted in an
almost pure $\gamma \gamma$ sample, where the contamination from 
Bhabha and Compton events is about 0.3\% 
and 3\% 
respectively.

\begin{figure}[hbt]
\begin{center}
 \mbox{\epsfig{file=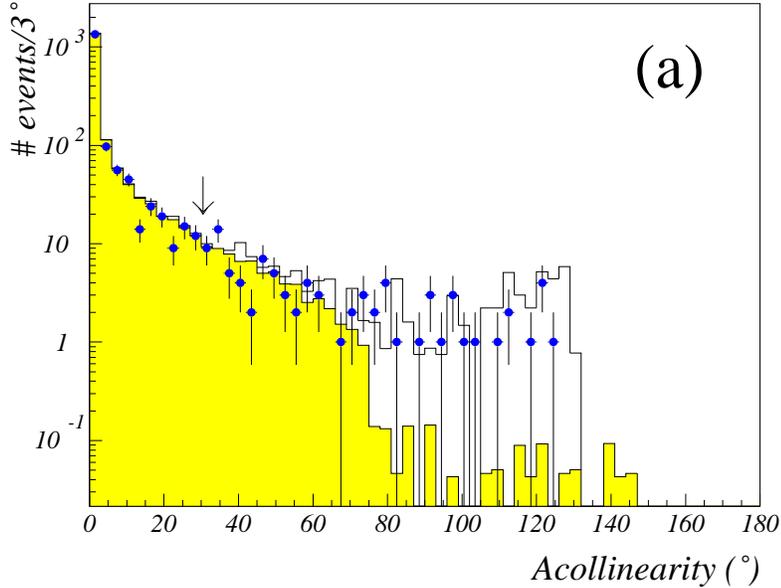,width=.7\linewidth}}
\end{center}
\caption{
Acollinearity distribution for the $\gamma \gamma (\gamma)$ sample
selected at all centre-of-mass energies (dots), 
before imposing the 30\gr\ acollinearity cut (arrow).
The histograms re\-pre\-sent the QED \eeintogg(\fot) simulation 
(grey area) and the remaining background (white area).
The latter is mainly due to Compton ($e^\pm\gamma$) events.}
\label{fig_acol}
\end{figure}

The radiation of a third hard photon constrains the
two harder photons to be produced at effective $\sqrt{s}$  
values which have been tested more accurately using lower energy data.    
Since the aim of this analysis is to test the QED 
$e^+e^- \to \gamma \gamma$ reaction at the highest available energies, 
such final states were not allowed in the selected sample:
events with a third hard {\it bremsstrahlung} photon can be considered as  
a higher order contribution to \eeintogg\ 
(like the soft {\it bremsstrahlung} and the virtual 
contributions), which can be deconvoluted from data by applying a radiative
correction factor when evaluating the $e^+e^- \to \gamma \gamma$ 
Born cross-section. Moreover, the $e^+e^- \to \gamma \gamma \gamma$ 
contribution can be dramatically reduced if the spatial angle 
between the two most energetic photons is required to be large.
Therefore, a final selection criterion, consisting in requiring 
that the acollinearity\footnote
{The acollinearity between
two directions is the complement to $\pi$ of the
spatial angle between them.} 
between the two most energetic photons was
below 30$^{\circ}$, was applied, 
eliminating most events with a third visible hard photon, 
and reducing the Compton background to 0.3\%. 
The acollinearity distribution prior to the cut  
is shown in figure \ref{fig_acol}(a) for the full data sample, and
compared to the \eeintogg(\fot)\ simulation and to the remaining 
background expectations.
After imposing all selection criteria, 
the contamination from Bhabha and Compton events to the selected  
$\gamma \gamma$ sample was estimated to be 0.6\%,
and taken into account in the systematic uncertainty.


\subsection{{\bf $\gamma \gamma$} trigger and selection efficiencies}

The trigger efficiency for neutral two-photon final states 
was computed with Bhabha events using the redundancy of the 
electromagnetic trigger with the track trigger.
It was calculated for each centre-of-mass energy 
as a function of $\rm |\cos{\theta^\ast}|$. The 
global values obtained for the barrel and endcaps 
are displayed in table \ref{tab_eft}.

Final states with one converted photon are triggered
by the single track coincidence trigger, whose efficiency 
is known to be near 100\%.
Two dedicated samples of Compton ($e^\pm \gamma$) events, 
one with a triggered FEMC photon and another with a triggered HPC photon, 
were used to cross-check the track trigger efficiency in the barrel and 
endcaps. The global efficiency for triggering events with one converted 
photon was confirmed to be above 99\% 
in both regions of the detector, for all data sets, and the 
resulting uncertainty was taken into account in the global 
systematic uncertainty.   

The selection efficiency for the two-photon event 
sample was evaluated as a function of $\rm |\cos{\theta^\ast}|$
using events from the 
\eeintogg(\fot) generator of Berends and Kleiss \cite{bib_BK} 
passed through the full DELPHI simulation and reconstruction chains 
\cite{bib_delphi}.
The effect of the calorimeter requirements on the 
selection efficiency obtained from simulation was cross-checked
using a sample of $e^+e^-$ events.
These events were selected using information coming exclusively from 
the tracking detectors.
The efficiency was defined as the ratio between the 
number of events in the subsample of 
$e^+e^-$ final states fulfilling the calorimetric selection 
and the total number of selected $e^+e^-$ events. This efficiency was 
computed as a function of $\rm |\cos{\theta^\ast}|$
for both real and simulated Bhabha events.
The difference observed between the efficiency for the data and for the 
simulation was taken as a systematic uncertainty in the \eeintogg(\fot) 
selection efficiency determination.

The global values for the selection efficiency, both in the 
barrel and in the forward region of DELPHI, are displayed in 
table \ref{tab_efs} along with their statistical and 
systematic uncertainties. 
A change in the forward DELPHI particle reconstruction algorithms
resulted in a better performance for $\gamma \gamma$ 
final states for the 1999 data processing compared with that of 1998.
However, there was an increase of the systematic 
uncertainty in the $\gamma \gamma$ selection efficiency.
  
\subsection{{\bf \eeintogg} cross-section}

The retained $\rm |cos\theta^\ast|$ acceptance was divided into 8 bins:
the barrel part of the detector, corresponding to
${\rm |\cos{\theta}^{\ast}| \in [0.035,0.731]}$ with 7 bins, (each covering ${\rm |\Delta \cos{\theta}^{\ast}| = 0.101}$,
except for the last bin, for which 
${\rm |\Delta \cos{\theta}^{\ast}| = 0.09}$)
and the forward region with one bin, 
${\rm |\cos{\theta}^{\ast}| \in[0.819,0.906]}$.
The number of events found in data for each centre-of-mass energy 
and the expected contribution from the QED process 
\eeintogg(\fot)\ 
(corrected for trigger efficiency) are displayed in table \ref{tab_nev} 
as a function of ${\rm |\cos{\theta}^{\ast}|}$.

The Born cross-section for the reaction $e^+ e^- \to \gamma \gamma (\gamma)$ 
was evaluated through expression (\ref{eq_stot}) for each centre-of-mass 
energy value,

\begin{equation}
{\rm \sigma^{0}_{dat}= \frac{N^{\gamma \gamma}}{{\cal{L}} \varepsilon R} \;\;[pb]}.
\label{eq_stot}
\end{equation}

\noindent 
N$^{\gamma \gamma}$ is the number of
selected events after background subtraction, 
$\cal{L}$ is the integrated luminosity,
$\varepsilon$ is the product of the  
selection and trigger efficiencies and R
is a radiative correction factor.
The radiative correction factor was
evaluated using the Monte Carlo generator of \cite{bib_BK}.
It was taken as the ratio between the $e^+ e^- \to \gamma \gamma (\gamma)$
cross-section computed up to order $\alpha^3$ to the Born 
cross-section (${\cal{O}}(\alpha^2)$) and found to be of the order of  
1.07 (1.04) for high (low) photon scattering angles.
 
A combined value of the Born cross-section at 
an average centre-of-mass energy of 193.8 GeV, 
corresponding to a total integrated luminosity of 375.7 pb$^{-1}$,
was obtained through expression (\ref{eq_stot}).
The average value of the centre-of-mass energy is obtained 
weighting the integrated luminosities of the different samples by 
the corresponding $s^{-1}$ factor.   

\begin{figure}[p]
\begin{center}
  \mbox{\epsfig{file=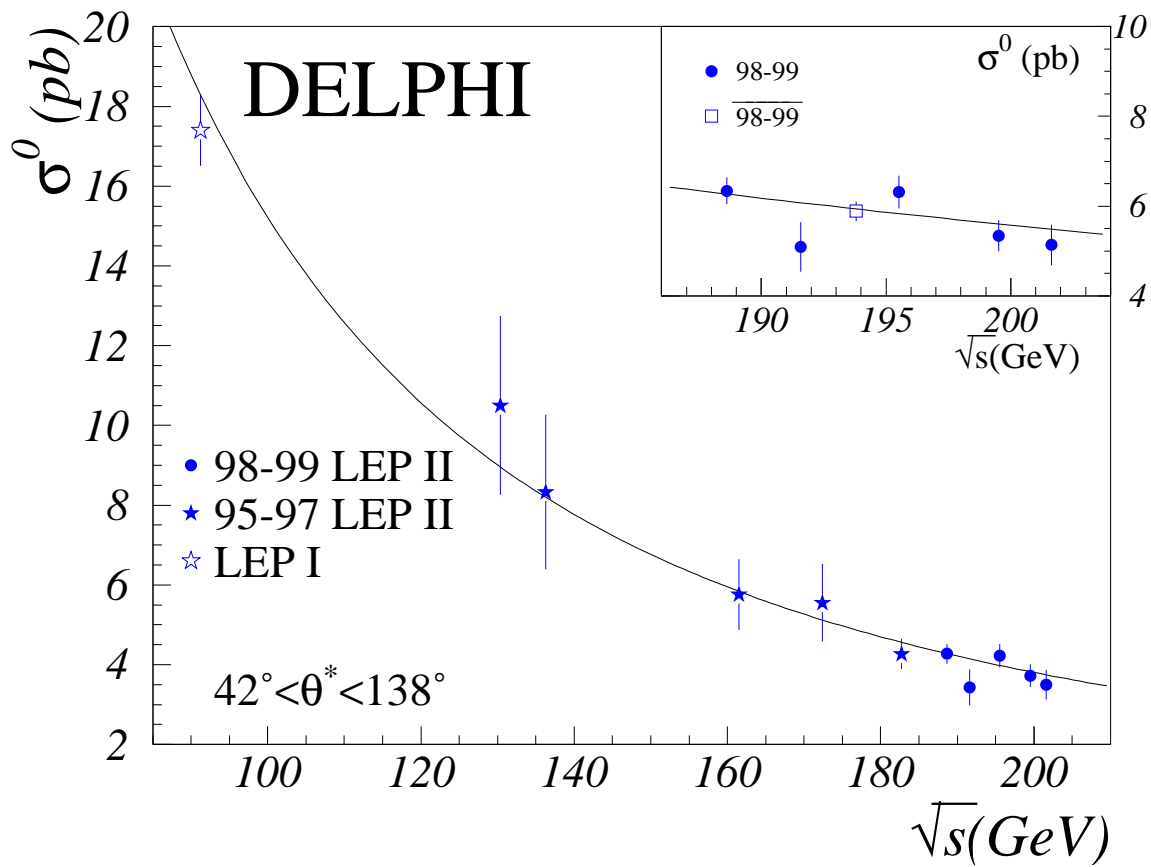,width=1.\linewidth}}
\end{center}
\caption{Born cross-section for \eeintogg\ 
in the barrel region of DELPHI,
$42^{\circ}<\theta^{\ast}<138^{\circ}$, as a function of the 
centre-of-mass energy, for 1990-1992 LEP I data (white star),
LEP II data collected between 1995 and 1997 (black stars),
and for the data collected during 1998 and 1999 (dots), 
compared to the QED prediction.
The Born cross-section measured within the analysis acceptance region
for the real data collected during 1998 and 1999 (dots) and the cross-section 
resulting from the combination of these data sets 
at an average centre-of-mass energy of 193.8 GeV 
(square) are compared to the QED prediction in the upper right plot.}
\label{fig_stot}
\end{figure}

$\rm N^{\gamma \gamma}$ is taken as the total number of selected events 
in the five data samples. 
The average trigger and selection efficiencies 
were obtained by weighting the global trigger and selection 
efficiencies of each data set 
by the corresponding integrated luminosities.
The measured Born cross-section for each of the five centre-of-mass 
energies and the combined result are compared to the QED 
predictions in table \ref{tab_stot} and in the upper right corner of
figure \ref{fig_stot}.
The $\chi^2$ of the measured values for the cross-section for 
the different centre-of-mass energies  
with respect to the QED prediction was  5.5 with 5 degrees of freedom. 

The Born cross-section values for the five centre-of-mass 
energies measured 
in the region $0.035 < |\cos{\theta^\ast}| < 0.731$, 
were corrected to the full barrel acceptance of DELPHI, 
$0.000 < |\cos{\theta^\ast}| < 0.742$, 
and the obtained values are presented in table \ref{tab_stot}. 
These are also displayed in figure \ref{fig_stot} as a function of 
the centre-of-mass energy, 
along with the previously published results, which include 
LEP I data collected between 1990 and 1992 \cite{bib_94} 
and former LEP II data collected between 1995 and 1997 \cite{bib_97}. 

The total systematic errors were obtained by adding in quadrature 
the uncertainties on the selection efficiency, trigger efficiencies,  
residual background, luminosity determination and on the 
radiative corrections (amounting to $\pm$0.5\%).
The systematic uncertainty in the selection efficiency
determination is the dominant contribution to the systematic error; 
with a typical value of $\pm$2.5\%. 
This uncertainty reflects residual differences between the real 
detector response and the simulated one. It is due to effects that cannot be 
fully described by the detector simulation such as detector instabilities 
and edge effects of calorimeters.
The uncertainty in the luminosity determination was $\pm$0.56\%. 
It was obtained by adding in quadrature the $\pm$0.5\% 
systematic uncertainty in the luminosity measurement and the 
$\pm$0.25\%
theoretical error in the Bhabha cross-section determination 
\cite{bib_lumi}. 

The \eeintogg\ differential Born cross-section was computed as:

\begin{equation}
{\rm \frac{d\sigma^0_i}{d\Omega}=
\frac{\sigma^0_i}{2 \pi \Delta \cos{\theta}^{\ast}_i} \;\; [pb/str],}
\label{eq_sdif}
\end{equation}

\noindent where ${\rm \sigma^0_i}$ stands for the measured Born
cross-section in each 
$\rm |\cos{\theta_{i}^{\ast}}|$ interval, (i).

\begin{figure}[p]
\begin{center}
  \mbox{\epsfig{file=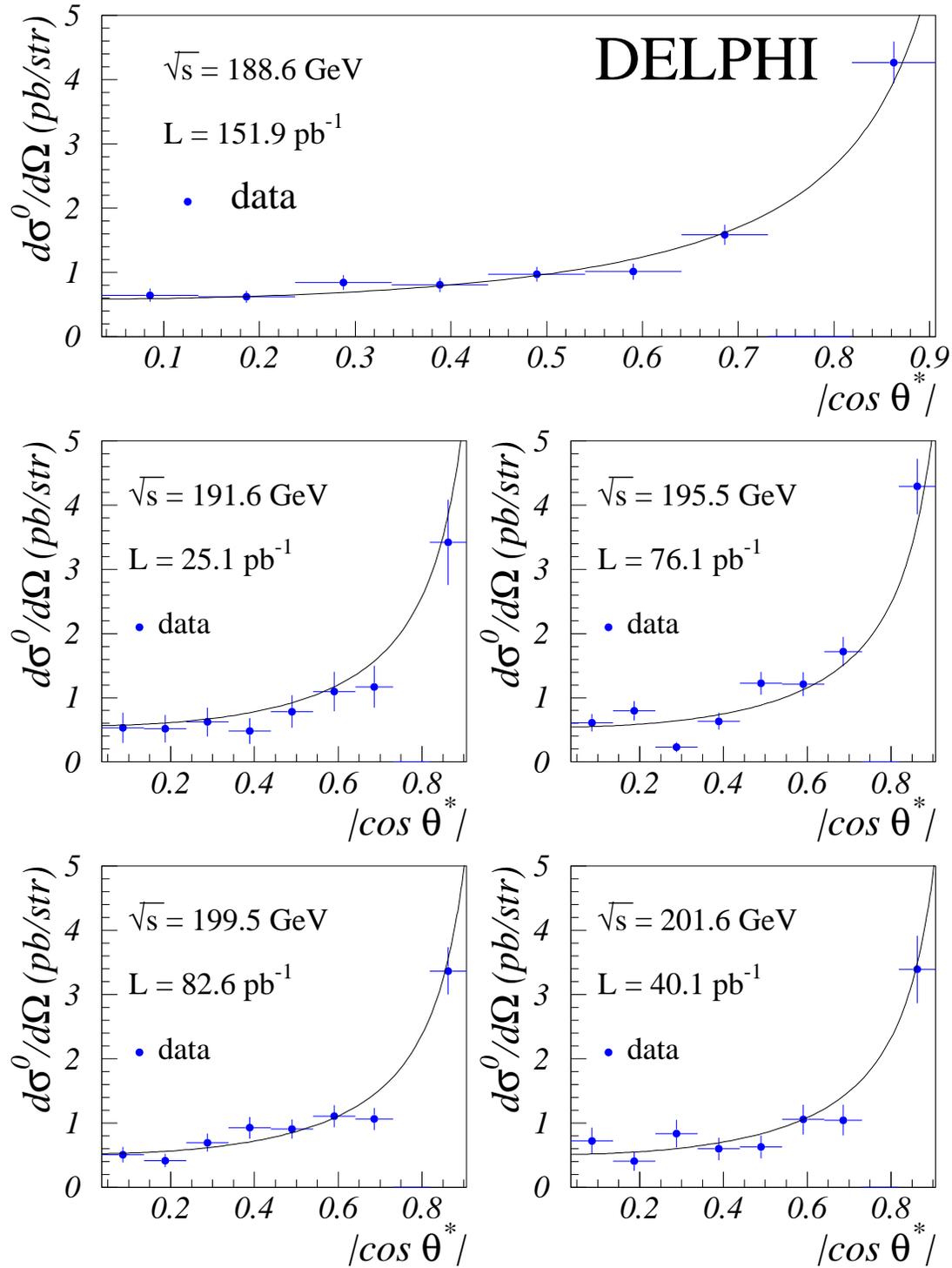,width=1.\linewidth}}
\end{center}
\caption{Differential Born cross-section distributions
obtained for the five centre-of-mass energies
compared to the corresponding QED theoretical predictions.}
\label{fig_dsdo_i}
\end{figure}

The differential cross-section was computed for each 
centre-of-mass energy, taking into account 
the $|\cos{\theta}^{\ast}|$ dependence of trigger and selection 
efficiencies, radiative corrections and their respective uncertainties.
Comparisons between the measured and predicted Born 
differential cross-sections for each centre-of-mass energy
are shown in figure \ref{fig_dsdo_i}.
The deficit of $\gamma \gamma$ events for 
$\rm |\cos{\theta}^{\ast}|$ 
between 0.237 and 0.338 for $\sqrt{s}$=195.5 GeV 
was concluded to be a statistical fluctuation: 
the trigger efficiency for this region was estimated to be about 98\% 
and the counting of energy deposits associated to Bhabha electrons in the
same $\rm |\cos{\theta}^{\ast}|$ region showed a good agreement 
with the simulation expectations.

The differential cross-section extracted from the combined 
data sets (corresponding to 
$\sqrt{s_{eff}}$~=~193.8~GeV),
is compared to the QED prediction in table \ref{tab_dsdo} and 
in figure \ref{fig_dsdo}.  
The $\chi^2$ of the differential cross-section binned distribution
at the mean centre-of-mass energy with respect to the 
QED prediction was 3.6 with 8 degrees of freedom.

\subsection{Deviations from QED}

Possible deviations from QED are described in the context of several models, 
which express the Born differential cross-section for \eeintogg\
as the sum of the QED term and of a deviation term:

\begin{equation}
{\rm \frac{d\sigma^0}{d\Omega}= 
\frac{\alpha^2}{s}\frac{1+cos^2\theta^\ast}{1-cos^2\theta^\ast} + 
\biggl( \frac{d\sigma}{d\Omega} \biggr)^D \;\;.}
\label{eq_dsqed}
\end{equation}

Among the models predicting deviations from QED are those described in
table \ref{tab_lim}. The most general parameterization consists of
introducing a cut-off parameter in the electron propagators ($\Lambda$), 
reflecting the energy scale up to which the e$\gamma$ interaction can be described as point-like \cite{bib_drell,bib_low}.

Deviations from QED could also follow from the 
t-channel exchange of an excited electron,
which, in composite models \cite{bib_composite}, is parameterized 
as a function of ${\rm \lambda_{\gamma}/M_{e^\ast}^2}$ 
(the ratio between the coupling of the excited electron 
to the photon and to the electron and the excited electron mass)
and of a kinematic factor, ${\rm H(cos^2\theta^{\ast})}$, 

\begin{equation}
{\rm H(cos^{2}\theta^\ast) = \frac{2M_{e^\ast}^2}{s} \cdot 
\biggl(\frac{2 M_{e^\ast}^2}{s}  +  
\frac{1-cos^{2}\theta^\ast}{1+cos^{2}\theta^\ast}\biggr) \huge/    
     \biggl[\biggl(1 + \frac{2M_{e^\ast}^2}{s} \biggr)^2 - 
     cos^{2}\theta^\ast \biggr].}
\label{eq_hmex}
\end{equation}

Deviations from the QED \eeintogg\ cross-section 
due to s-channel exchange of virtual gravitons 
were also probed.
These can be parameterized as a function of $\rm \frac{\lambda}{M_s^4}$,
where $\rm M_s$ is the string mass scale, 
which in some string models could be of the order of the electroweak scale
\cite{bib_giudice,bib_agashe}. 
$\rm \lambda$ is a parameter entering Quantum Gravity models, 
conventionally taken to be $\pm$ 1.
The ratio $\rm \frac{\lambda}{M_s^4}$ follows the notation of \cite{bib_hewett}
and is related to the Quantum Gravity 
scale, $\rm \Lambda_T$, in reference \cite{bib_giudice} via:

\begin{equation}
{\rm \frac{|\lambda|}{M_s^4} = 
\frac{\pi}{2}\frac{1}{\Lambda_T^4}\;\; [GeV^{-4}].}
\label{eq_ms}
\end{equation}

\begin{figure}[p]
\begin{center}
  \mbox{\epsfig{file=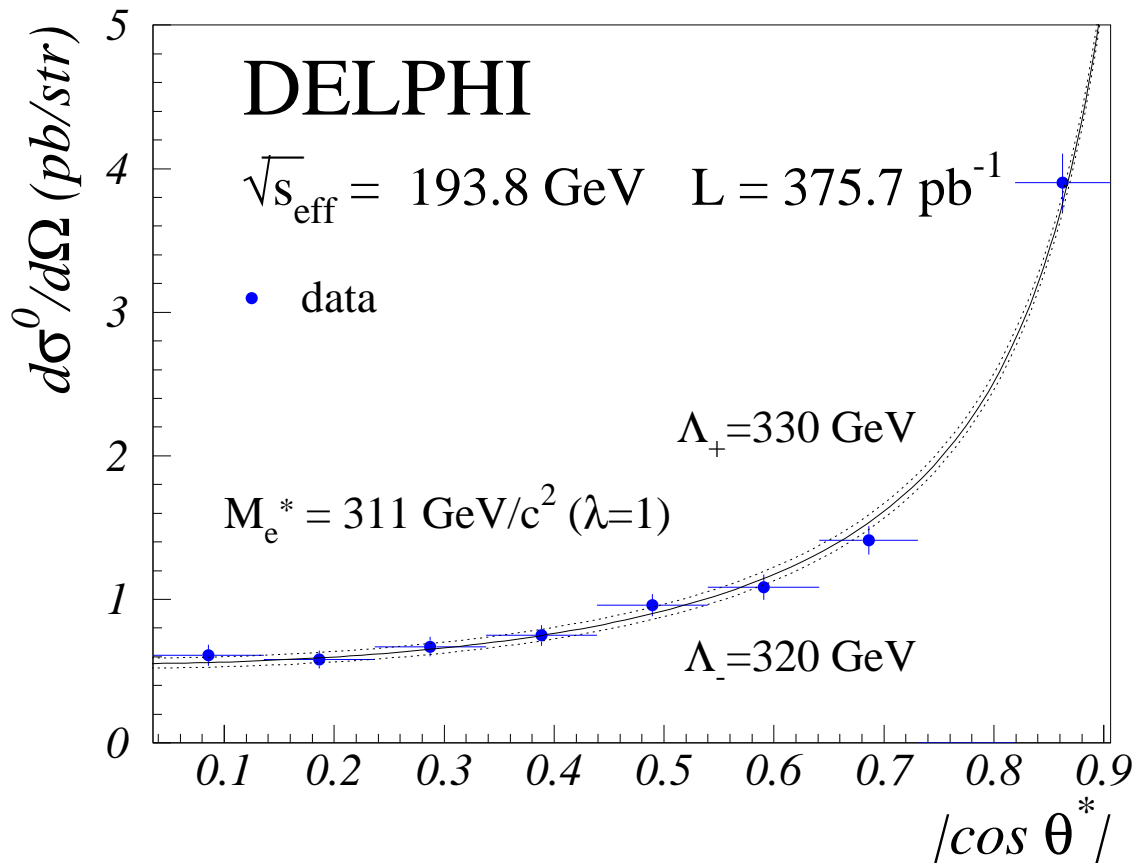,width=1.\linewidth}}
\end{center}
\caption{Born differential cross-section obtained by combining all 
data sets at an effective centre-of-mass energy of 193.8 GeV (dots),
compared to the QED theoretical distribution (full line). 
The dotted lines represent the allowed 95\% 
C.L. deviations from the QED differential cross-section, 
which correspond to 95\% 
C.L. lower limits on $\Lambda_+$ and $\Lambda_-$ of 330 GeV 
and 320 GeV respectively, to a 311 GeV/c$^2$ 95\% 
C.L. lower limit 
on the excited electron mass (for $\lambda_\gamma$ = 1 ), and to 95\% 
C.L. lower limits on the string mass scale of 
713 GeV/c$^2$ (for  $\lambda$ = 1 ) and  
691 GeV/c$^2$ (for  $\lambda$ = -1 ).}
\label{fig_dsdo}
\end{figure}

The 95\% 
C.L. limits were extracted for the free parameters in these models.
This was achieved using a binned maximum likelihood function, by
renormalizing the joint probability to the physical region of each parameter 
according to the Bayesian approach described in \cite{bib_pdg}. 
The cross-section parameterization for the models considered, the chosen 
estimators ($\rm \xi$) and the results of the likelihood function 
maximization are displayed in table \ref{tab_lim} along with the 95\% 
C.L. lower limits on each model parameter, $\rm \Lambda$, $\rm M_e^{\ast}$ 
and $\rm M_s$.
The changes in the differential cross-section resulting from the 
range of fitted parameters are indicated by the dotted lines 
in figure \ref{fig_dsdo}. 
The final results presented in table \ref{tab_lim} and in figure \ref{fig_dsdo} 
were obtained by combining the results of 
the present analysis with results published previously 
\cite{bib_97}.
The latter are based 
on LEP I data taken between 1990 and 1992, 
and on LEP II data collected between 1995 and 1997.
Their centre-of-mass energies range from 91.2 GeV up to 182.7 GeV,
corresponding to an integrated luminosity of 115.1 pb$^{-1}$.

\section{Summary}

The reaction \eeintogg(\fot) was studied using the LEP 1998 and 1999 
high energy data, collected with the DELPHI detector at centre-of-mass 
energies of 188.6 GeV, 191.6 GeV, 195.5 GeV, 199.5 GeV and 201.6 GeV,
corresponding to integrated luminosities of 151.9 pb$^{-1}$, 
25.1 pb$^{-1}$, 76.1 pb$^{-1}$, 82.6 pb$^{-1}$ and 40.1 pb$^-1$ respectively.
The differential and total cross-sections for the process 
$e^+e^- \to \gamma \gamma$ were measured. 
Good agreement between the data and the QED prediction for this process 
was found. Lower limits on possible deviations from QED were derived by 
combining the present analysis result with a previously published one 
\cite{bib_97}. The 95\% 
C.L. lower limits on the QED cut-off parameters 
of $\rm \Lambda_{+}>$330 GeV and $\rm \Lambda_{-}>$ 320 GeV 
were obtained. In the framework of composite models, a 95\% 
C.L. lower limit for the mass of an excited electron, 
${\rm M_{e^{\ast}} >}$ 311 GeV/c$^2$, was obtained
considering an effective coupling value of 1 for $\rm \lambda_{\gamma}$.
The  possible contribution of virtual gravitons to the process 
$e^+e^- \to \gamma \gamma$ was probed, resulting in 95\% 
C.L. lower limits in the string mass scale of 
$\rm M_S > 713 \, GeV/c^2$ and $\rm M_S > 691 \,GeV/c^2$ 
for $\rm \lambda=1$ and $\rm \lambda=-1$ respectively (where $\rm \lambda$ 
is a ${\cal{O}}(1)$ parameter of Quantum Gravity models).

\subsection*{Acknowledgements}

 We are greatly indebted to our technical 
collaborators, to the members of the CERN-SL Division for the excellent 
performance of the LEP collider, and to the funding agencies for their
support in building and operating the DELPHI detector.\\
We acknowledge in particular the support of \\
Austrian Federal Ministry of Science and Traffics, GZ 616.364/2-III/2a/98, \\
FNRS--FWO, Belgium,  \\
FINEP, CNPq, CAPES, FUJB and FAPERJ, Brazil, \\
Czech Ministry of Industry and Trade, GA CR 202/96/0450 and GA AVCR A1010521,\\
Danish Natural Research Council, \\
Commission of the European Communities (DG XII), \\
Direction des Sciences de la Mati\`ere, CEA, France, \\
Bundesministerium f$\ddot{\mbox{\rm u}}$r Bildung, Wissenschaft, Forschung 
und Technologie, Germany,\\
General Secretariat for Research and Technology, Greece, \\
National Science Foundation (NWO) and Foundation for Research on Matter (FOM),
The Netherlands, \\
Norwegian Research Council,  \\
State Committee for Scientific Research, Poland, 2P03B06015, 2P03B03311 and
SPUB/P03/178/98, \\
JNICT--Junta Nacional de Investiga\c{c}\~{a}o Cient\'{\i}fica 
e Tecnol\'ogica, Portugal, \\
Vedecka grantova agentura MS SR, Slovakia, Nr. 95/5195/134, \\
Ministry of Science and Technology of the Republic of Slovenia, \\
CICYT, Spain, AEN96--1661 and AEN96-1681,  \\
The Swedish Natural Science Research Council,      \\
Particle Physics and Astronomy Research Council, UK, \\
Department of Energy, USA, DE--FG02--94ER40817. \\


\pagebreak

{\large 
\begin{table}[h]
\vspace{3cm}
\begin{center}
\begin{tabular}{|c|c|c|}
\hline
\multicolumn{1}{|c|}{ }           &
\multicolumn{2}{c|}{$\epsilon_{trigger}^{\gamma \gamma}$} \\
\cline{2-3}
\multicolumn{1}{|c|}{ $\sqrt{s}$}  & 
\multicolumn{1}{c|}{ Barrel}      &
\multicolumn{1}{c|}{ Forward}     \\
\multicolumn{1}{|c|}{[GeV]}   & 
\multicolumn{1}{|c|}{ $\rm |cos{\theta^{\ast}}| \in [0.035,0.731]$} & 
\multicolumn{1}{|c|}{ $\rm |cos{\theta^{\ast}}| \in [0.819,0.906]$} \\ 
\cline{2-3}
\hline
188.6 & 0.985 $\pm$ 0.002 &  1.0000 $\pm$ 0.0003     \\
191.6 & 0.977 $\pm$ 0.007 &  1.000 $\pm$ 0.002     \\
195.5 & 0.977 $\pm$ 0.004 &  0.9995 $\pm$ 0.0005   \\
199.5 & 0.968 $\pm$ 0.005 &  0.9995 $\pm$ 0.0005   \\
201.6 & 0.983 $\pm$ 0.005 &  1.000 $\pm$ 0.001     \\
\hline
\end{tabular}
\end{center}
\vspace{0.5cm}
\caption{ Trigger efficiencies (using the redundancy of the 
Bhabha trigger) for $\gamma \gamma$ 
neutral final states in the barrel and forward regions of the 
detector for the different data sets.}
\label{tab_eft}
\end{table}
}

{\large	
\begin{table}[h]
\vspace{2cm}
\begin{center}
\begin{tabular}{|c|c|c|}
\hline
\multicolumn{1}{|c|}{ }                                      & 
\multicolumn{2}{c|}{$\epsilon_{sel}^{\gamma \gamma + \gamma \gamma_c}$} \\
\cline{2-3}
\multicolumn{1}{|c|}{$\sqrt{s}$}  & 
\multicolumn{1}{c|}{ Barrel } & 
\multicolumn{1}{c|}{ Forward } \\ 
\multicolumn{1}{|c|}{[GeV]}                                      & 
\multicolumn{1}{c|}{ $\rm |cos\theta^\ast| \in [0.035,0.731]$} &
\multicolumn{1}{c|}{ $\rm |cos\theta^\ast| \in [0.819,0.906]$} \\
\hline
188.6       & 0.754 $\pm$ 0.004 $\pm$ 0.032 
& 0.480 $\pm$ 0.006 $\pm$ 0.003\\
191.6 - 201.6 & 0.756 $\pm$ 0.004 $\pm$ 0.029 
& 0.557 $\pm$ 0.007 $\pm$ 0.012 \\
\hline
\end{tabular}
\end{center}
\vspace{0.5cm}
\caption{
Selection efficiencies for  $\gamma \gamma (\gamma)$ final states
in the barrel and forward regions of the detector,
with their statistical and systematic uncertainties, 
for the two data taking periods.}
\label{tab_efs}
\end{table} 
}

\begin{table}[p]
{\normalsize
\begin{center}
\setlength{\tabcolsep}{1.5mm}
\begin{tabular}{|l|l|rl|rl|r|}
\hline
& \multicolumn{1}{|c|}{\scalebox{1}[1.3]{$\rm |\cos{\theta}^{\ast} |$ } } & 
\multicolumn{1}{|r}{ $\rm N^{\gamma \gamma+\gamma\gamma_c}_{dat}$ }    & 
\multicolumn{1}{l|}{ ($\rm N_{QED}\pm \Delta N_{stat}$)} &
\multicolumn{1}{|l}{ $\rm N^{\gamma \gamma_c}_{dat}$ }    & 
\multicolumn{1}{r|}{ ($\rm N_{QED}$)} &
\multicolumn{1}{|c|}{ $\rm d\sigma^{0}_{dat}/d\Omega$ [pb/str]}   \\ 
\cline{2-7}
&0.035-0.136 & 46 & (41.5 $\pm$ 1.4) &5 &(6.2) & 0.65 $\pm$ 0.10 $\pm$ 0.05\\
&0.136-0.237 & 48 & (47.9 $\pm$ 1.5) &3 &(3.8) & 0.62 $\pm$ 0.09 $\pm$ 0.01\\
&0.237-0.338& 64 & (52.6 $\pm$ 1.6) &5 &(6.4) & 0.84 $\pm$ 0.11 $\pm$ 0.04\\
&0.338-0.439& 57 & (54.8 $\pm$ 1.5) &5 &(6.2) & 0.81 $\pm$ 0.11 $\pm$ 0.03\\
&0.439-0.540& 77 & (71.1 $\pm$ 1.8) &11 &(8.5) & 0.97 $\pm$ 0.11 $\pm$ 0.03\\
&0.540-0.641& 76 & (90.0 $\pm$ 2.0) &19 &(10.8) & 1.01 $\pm$ 0.12 $\pm$ 0.04\\
&0.641-0.731& 108& (111.7 $\pm$ 2.3)&11 &(15.8) & 1.59 $\pm$ 0.15 $\pm$ 0.04\\
&0.819-0.906& 176& (170.3 $\pm$ 2.8)& 47 & (53.4) &4.27 $\pm$ 0.32 $\pm$ 0.06\\
\cline{2-7}
\raisebox{8mm}[0mm][0mm]{%
\rotatebox{90}{%
$\sqrt{s}$ = 188.6 GeV}}%
& total &652 & (639.7 $\pm$ 5.4)&106&(111.1 )& \\
\hline
\hline
& \multicolumn{1}{|c|}{\scalebox{1}[1.3]{$\rm |\cos{\theta}^{\ast} |$ } } & 
\multicolumn{1}{|r}{ $\rm N^{\gamma \gamma + \gamma \gamma_c}_{dat}$ }    & 
\multicolumn{1}{l|}{ ($\rm N_{QED}\pm \Delta N_{stat}$)} &
\multicolumn{1}{|l}{ $\rm N^{\gamma \gamma_c}_{dat}$ }    & 
\multicolumn{1}{r|}{ ($\rm N_{QED}$)} &
\multicolumn{1}{|c|}{ $\rm d\sigma^{0}_{dat}/d\Omega$ [pb/str]}   \\ 
\cline{2-7}
&0.035-0.136 & 6 &(6.4$\pm$0.3) & 2 &(0.9 )& 0.53 $\pm$ 0.22 $\pm$ 0.09\\
&0.136-0.237 & 6 &(7.2$\pm$0.3) & 0 &(0.7 )& 0.52 $\pm$ 0.21 $\pm$ 0.05\\
&0.237-0.338& 8 &(8.5$\pm$0.3)  & 1 &(0.9 )& 0.62 $\pm$ 0.22 $\pm$ 0.04\\
&0.338-0.439& 6 &(9.9$\pm$0.3)  & 1 &(1.1 )& 0.48 $\pm$ 0.20 $\pm$ 0.03\\
&0.439-0.540& 10 &(12.4$\pm$0.4)& 1 &(1.5 )& 0.79 $\pm$ 0.25 $\pm$ 0.05\\
&0.540-0.641& 14 &(14.7$\pm$0.4)& 5 & (1.8)& 1.09 $\pm$ 0.29 $\pm$ 0.09\\
&0.641-0.731& 13 &(17.8$\pm$0.4)& 1 & (2.7)& 1.17 $\pm$ 0.32 $\pm$ 0.03\\
&0.819-0.906& 27 &(31.3$\pm$0.6)& 8 & (7.7)& 3.42 $\pm$ 0.66 $\pm$ 0.09\\
\cline{2-7}
\raisebox{8mm}[0mm][0mm]{%
\rotatebox{90}{%
$\sqrt{s}$ = 191.6 GeV}}%
&total&90&(108.2$\pm$1.1)& 19& (17.3 )& \\
\hline
\hline
& \multicolumn{1}{|c|}{\scalebox{1}[1.3]{$\rm |\cos{\theta}^{\ast} |$ } } & 
\multicolumn{1}{|r}{ $\rm N^{\gamma \gamma + \gamma \gamma_c}_{dat}$ }    & 
\multicolumn{1}{l|}{ ($\rm N_{QED}\pm \Delta N_{stat}$)} &
\multicolumn{1}{|l}{ $\rm N^{\gamma \gamma_c}_{dat}$ }    & 
\multicolumn{1}{r|}{ ($\rm N_{QED}$)} &
\multicolumn{1}{|c|}{ $\rm d\sigma^{0}_{dat}/d\Omega$ [pb/str]}   \\ 
\cline{2-7}
&0.035-0.136 & 21 &(19.3$\pm$0.8)& 4 &(2.5 )& 0.61 $\pm$ 0.13 $\pm$ 0.03\\
&0.136-0.237 & 29 &(21.5$\pm$0.8)& 5 &(2.1 )& 0.80 $\pm$ 0.15 $\pm$ 0.04\\
&0.237-0.338 & 9  &(24.6$\pm$0.9)& 0 &(2.5 )& 0.23 $\pm$ 0.08 $\pm$ 0.01\\
&0.338-0.439 & 23 &(27.4$\pm$0.9)& 4 &(3.2 )& 0.64 $\pm$ 0.13 $\pm$ 0.02\\
&0.439-0.540 & 48 &(36.6$\pm$1.1)& 2 &(4.4 )& 1.23 $\pm$ 0.18 $\pm$ 0.03\\
&0.540-0.641 & 47 &(43.2$\pm$1.2)& 6 &(5.3 )& 1.21 $\pm$ 0.18 $\pm$ 0.06\\
&0.641-0.731& 58 &(51.7$\pm$1.7) &12 &(7.9 )& 1.72 $\pm$ 0.23 $\pm$ 0.04\\
&0.819-0.906& 102 &(91.0$\pm$1.7)&28 &(22.6 )& 4.29 $\pm$ 0.42 $\pm$ 0.08\\
\cline{2-7}
\raisebox{8mm}[0mm][0mm]{%
\rotatebox{90}{%
$\sqrt{s}$ = 195.5 GeV}}%
&total& 337&(315.3 $\pm$ 3.1)&61 &(50.5 ) &\\
\hline
\hline
& \multicolumn{1}{|c|}{\scalebox{1}[1.3]{$\rm |\cos{\theta}^{\ast} |$ } } & 
\multicolumn{1}{|r}{ $\rm N^{\gamma \gamma+\gamma\gamma_c}_{dat}$ }    & 
\multicolumn{1}{l|}{ ($\rm N_{QED}\pm \Delta N_{stat}$)} &
\multicolumn{1}{|l}{ $\rm N^{\gamma \gamma_c}_{dat}$ }    & 
\multicolumn{1}{r|}{ ($\rm N_{QED}$)} &
\multicolumn{1}{|c|}{ $\rm d\sigma^{0}_{dat}/d\Omega$ [pb/str]}   \\ 
\cline{2-7}
&0.035-0.136 & 19 &(21.2$\pm$0.8)& 3 &(2.5)& 0.51 $\pm$ 0.12 $\pm$ 0.03 \\
&0.136-0.237 & 17 &(23.0$\pm$0.9)& 0 &(2.5 )& 0.42 $\pm$ 0.10 $\pm$ 0.03\\
&0.237-0.338 & 28 &(23.7$\pm$0.9)& 2 &(2.4 )& 0.70 $\pm$ 0.13 $\pm$ 0.04\\
&0.338-0.439 & 34 &(25.0$\pm$0.9)& 3 &(3.7 )& 0.93 $\pm$ 0.16 $\pm$ 0.05\\
&0.439-0.540 & 39 &(35.2$\pm$1.1)& 1 &(4.5 )& 0.91 $\pm$ 0.15 $\pm$ 0.03\\
&0.540-0.641 & 45 &(45.4$\pm$1.2)& 4 &(6.3 )& 1.11 $\pm$ 0.16 $\pm$ 0.05\\
&0.641-0.731& 40 &(54.9$\pm$1.4) & 9 &(6.8 )& 1.07 $\pm$ 0.17 $\pm$ 0.03\\
&0.819-0.906& 88 &(96.0$\pm$1.8) &29 &(24.4 )& 3.37 $\pm$ 0.36 $\pm$ 0.09\\
\cline{2-7}
\raisebox{8mm}[0mm][0mm]{%
\rotatebox{90}{%
$\sqrt{s}$ = 199.5 GeV}}%
&total& 310&(324.3$\pm$3.3)&51 & (53.1)& \\
\hline
\hline
& \multicolumn{1}{|c|}{\scalebox{1}[1.3]{$\rm |\cos{\theta}^{\ast} |$ } } & 
\multicolumn{1}{|r}{ $\rm N^{\gamma \gamma+\gamma\gamma_c}_{dat}$ }    & 
\multicolumn{1}{l|}{ ($\rm N_{QED}\pm \Delta N_{stat}$)} &
\multicolumn{1}{|l}{ $\rm N^{\gamma \gamma_c}_{dat}$ }    & 
\multicolumn{1}{r|}{ ($\rm N_{QED}$)} &
\multicolumn{1}{|c|}{ $\rm d\sigma^{0}_{dat}/d\Omega$ [pb/str]}   \\ 
\cline{2-7}
&0.035-0.136 & 14 &(10.7$\pm$0.4)& 2 &(1.2 )& 0.72 $\pm$ 0.19 $\pm$  0.06\\
&0.136-0.237 & 8 &(10.9$\pm$0.4) & 0 &(1.2)& 0.40 $\pm$ 0.14 $\pm$ 0.04\\
&0.237-0.338& 17 &(11.7$\pm$0.4) & 4 &(1.2 )& 0.84 $\pm$ 0.20 $\pm$ 0.08 \\
&0.338-0.439 & 12 &(13.3$\pm$0.5)& 1 &(1.8 )& 0.60 $\pm$ 0.17 $\pm$ 0.03\\
&0.439-0.540& 13&(16.6$\pm$0.5)  & 0 &(2.1)& 0.63 $\pm$ 0.17 $\pm$ 0.03 \\
&0.540-0.641 & 21&(21.7$\pm$0.6) & 1 &(3.0 )& 1.06 $\pm$ 0.23 $\pm$ 0.04\\
&0.641-0.731& 19&(26.1$\pm$0.6)  & 4 &(3.2 )& 1.05 $\pm$ 0.24 $\pm$ 0.03\\
&0.819-0.906& 43&(45.6$\pm$0.9) & 15 &(11.6 )& 3.39 $\pm$ 0.52 $\pm$ 0.10\\
\cline{2-7}
\raisebox{8mm}[0mm][0mm]{%
\rotatebox{90}{%
$\sqrt{s}$ = 201.6 GeV}}%
&total& 147 &(156.6$\pm$1.6)& 27 &(25.3 )& \\
\hline
\end{tabular}
\end{center}
}
\caption{ Number of events selected from data as a function of 
$\rm |\cos{\theta^\ast}|$ and number of expected events (in parenthesis) 
from QED corrected for trigger efficiency. 
The uncertainties associated 
to the QED predictions are statistical only.
In the third column the number of events with one converted 
photon is given along with the QED si\-mu\-la\-tion prediction.
In the fourth column the measured Born differential cross-section
is displayed with statistical and systematic uncertainties.
\label{tab_nev} }
\end{table}

\begin{table}[h]
\begin{center}
\begin{tabular}{|c|c|c|c|c|}
\hline
\multicolumn{1}{|c|}{$\sqrt{s}$} &  
\multicolumn{2}{c|}{analysis acceptance} & 
\multicolumn{2}{c|}{$\cos{\theta^\ast} \in [-0.742,0.742]$} \\
\cline{2-5}
\multicolumn{1}{|c|}{[GeV]} &  
\multicolumn{1}{c|}{$\sigma^0_{dat}$ [pb]} & 
\multicolumn{1}{c|}{$\sigma^0_{QED}$ [pb]} &
\multicolumn{1}{c|}{$\sigma^0_{dat}$ [pb]} & 
\multicolumn{1}{c|}{$\sigma^0_{QED}$ [pb]}  \\
\hline
\hline
188.6  & 6.34 $\pm$ 0.25 $\pm$ 0.16    &  6.27 & 
4.27 $\pm$ 0.20 $\pm$ 0.14 & 4.28 \\
191.6  & 5.09 $\pm$ 0.54 $\pm$ 0.13    &  6.08 & 
3.43 $\pm$ 0.43 $\pm$ 0.11 & 4.15 \\
195.5  & 6.31 $\pm$ 0.34 $\pm$ 0.13    &  5.83 &
4.22 $\pm$ 0.28 $\pm$ 0.09 & 3.98 \\
199.5  & 5.34 $\pm$ 0.30 $\pm$ 0.17    &  5.60 &
3.73 $\pm$ 0.25 $\pm$ 0.14 & 3.82 \\
201.6  & 5.14 $\pm$ 0.42 $\pm$ 0.16    &  5.49 &
3.50 $\pm$ 0.34 $\pm$ 0.13 & 3.74 \\
\hline
193.8  & 5.89 $\pm$ 0.15 $\pm$ 0.16    &  5.94 & 
4.00 $\pm$ 0.12 $\pm$ 0.12 & 4.05 \\
\hline
\end{tabular}
\end{center}
\caption{ Measured Born cross-sections 
for \eeintogg\ 
(with statistical and systematic uncertainties) 
at the different centre-of-mass energies, for the analysis 
$\cos{\theta^\ast}$ acceptance and for the barrel region
($42^{\circ}<\theta^{\ast}<138^{\circ}$), compared to the  
corresponding QED predictions. In the last line the combined 
results are displayed along with the QED cross-sections 
at a centre-of-mass energy of 193.8 GeV.}
\label{tab_stot}
\end{table}

\begin{table}[h]
\begin{center}
\begin{tabular}{|c|c|c|}
\hline
\multicolumn{1}{|c|}{${\rm |\cos{ \theta^{\ast} }| }$}   & 
\multicolumn{1}{|c|}{${\rm d\sigma^0_{dat}/d\Omega}$ [pb/str]}       & 
\multicolumn{1}{|c|}{${\rm d\sigma^0_{QED}/d\Omega}$ [pb/str]}  \\  \hline
0.035-0.136 & 0.61 $\pm$ 0.06$\pm$ 0.04&  0.56 \\
0.136-0.237 & 0.58 $\pm$ 0.06$\pm$ 0.03&  0.59 \\
0.237-0.338 & 0.67 $\pm$ 0.06$\pm$ 0.03&  0.65 \\
0.338-0.439 & 0.75 $\pm$ 0.07$\pm$ 0.03&  0.75 \\
0.439-0.540 & 0.96 $\pm$ 0.07$\pm$ 0.03&  0.90 \\
0.540-0.641 & 1.08 $\pm$ 0.08$\pm$ 0.04&  1.14 \\
0.641-0.731 & 1.41 $\pm$ 0.09$\pm$ 0.03&  1.53 \\
0.819-0.906 & 3.90 $\pm$ 0.19$\pm$ 0.08&  3.76 \\
\hline
\end{tabular}
\end{center}
\caption{
Measured and predicted Born differential cross-section 
(the measured cross-section uncertainties are statistical and 
systematic) 
for the QED process \eeintogg\ at a mean centre-of-mass 
energy of 193.8 GeV obtained by combining 
the data sets corresponding to centre-of-mass energies
of 189.6 GeV, 191.6 GeV, 195.5 GeV, 199.5 GeV and 201.6 GeV.
 \label{tab_dsdo}}
\end{table}

{\normalsize
\begin{table}[hbt]
\begin{center}
\begin{tabular}{|c|c c|c|c c|}
\hline
\multicolumn{1}{|c|}{  }           &
\multicolumn{2}{|c|}{\raisebox{-3pt} {Cut-off}}           &
\multicolumn{1}{|c|}{\raisebox{-3pt}{Composite}}           &
\multicolumn{2}{|c|}{\raisebox{-3pt}{Low $M_s$}}           \\
& & & & & \\
\hline
\multicolumn{1}{|c|}{\raisebox{-4pt}
{$\rm (\frac{d\sigma}{d\Omega})^D$}} & 
\multicolumn{2}{|c|}{\raisebox{-4pt}
{$\rm \frac{\alpha^2 s}{2} (1+cos^2\theta^\ast)\cdot \xi $}}        &
\multicolumn{1}{|c|}{\raisebox{-4pt}
{$\rm \frac{\alpha^2 s}{2}(1+cos^2\theta^\ast) H(cos^2\theta^\ast)\cdot \xi$}}  &
\multicolumn{2}{|c|}{\raisebox{-4pt}{$\rm \frac{\alpha s}{4\pi} (1+cos^2\theta^\ast)\cdot\xi +{\cal{O}}(\xi^2)$}}           \\
& & & & & \\
\hline
\multicolumn{1}{|c|}{\raisebox{-3pt}{$\large \rm \xi $}} & 
\multicolumn{2}{|c|}{\raisebox{-3pt}
{$\rm  \pm 1/\Lambda^4_{\pm}$}}         &
\multicolumn{1}{|c|}{\raisebox{-3pt}{$\rm (\lambda_{\gamma}/M_{e^\ast}^2)^2$}}      &
\multicolumn{2}{|c|}{\raisebox{-3pt}{
$\rm \lambda/M_{s}^4$}} \\
& & & & & \\
\multicolumn{1}{|c|}{$\large \rm \xi^{+\sigma_+}_{-\sigma_-}$} &
\multicolumn{2}{|c|}{$ (0.034^{+0.547}_{-0.530})10^{-10}$} & 
\multicolumn{1}{|c|}{$(0.048^{+0.679}_{-0.611})10^{-10}$} &
\multicolumn{2}{|c|}{$(0.015^{+0.251}_{-0.243})10^{-11}$} \\
\raisebox{-2pt}{(1998-1999)} & & & & & \\
& & & & &\\
\multicolumn{1}{|c|}{$\rm \xi^{+\sigma_+}_{-\sigma_-}$} &
\multicolumn{2}{|c|}{$(-0.131^{+0.515}_{-0.501})10^{-10}$} & 
\multicolumn{1}{|c|}{$(-0.176^{+0.654}_{-0.599})10^{-10}$} & 
\multicolumn{2}{|c|}{$(-0.060^{+0.236}_{-0.230})10^{-11}$} \\ 
\raisebox{-2pt}{(1990-1999)} & & & & &\\
& & & & &\\
\hline
\multicolumn{1}{|c|}{\raisebox{-4pt} {95\% C.L.} } & 
\multicolumn{1}{|c|}{\raisebox{-2pt}{$\Lambda_+$ [GeV]}} & 
\multicolumn{1}{|c|}{\raisebox{-2pt}{$\Lambda_-$ [GeV]}} & 
\multicolumn{1}{|c|}{\raisebox{-2pt}{$M_{e^{\ast}}$ [GeV/c$^2$]}} & 
\multicolumn{1}{|c|}{\raisebox{-2pt}{$M_{s}$ [GeV/c$^2$]}}   & 
\multicolumn{1}{|c|}{\raisebox{-2pt}{$M_{s}$ [GeV/c$^2$]}}   \\
\multicolumn{1}{|c|}{ } & 
\multicolumn{1}{|c|}{ } & 
\multicolumn{1}{|c|}{ } & 
\multicolumn{1}{|c|}{$\lambda_\gamma=1$ } & 
\multicolumn{1}{|c|}{$\lambda= +1$ } & 
\multicolumn{1}{|c|}{$\lambda= -1$ } \\
\cline{2-6}
\multicolumn{1}{|c|}{\raisebox{2pt} {lower limits} } & 
\multicolumn{1}{|c|}{\raisebox{-4pt} {330}} & 
\multicolumn{1}{|c|}{\raisebox{-4pt} {320}} & 
\multicolumn{1}{|c|}{\raisebox{-4pt} {311}} & 
\multicolumn{1}{|c|}{\raisebox{-4pt} {713}} & 
\multicolumn{1}{|c|}{\raisebox{-4pt} {691}} \\
\multicolumn{1}{|c|}{ } &
\multicolumn{1}{|c|}{ } & 
\multicolumn{1}{|c|}{ } & 
\multicolumn{1}{|c|}{ } & 
\multicolumn{1}{|c|}{ } & 
\multicolumn{1}{|c|}{ } \\
\hline
\end{tabular}
\end{center}
\caption{Parameterization for each model predicting a deviation from QED, 
chosen estimator ($\rm \xi$), 
output of the likelihood function maximization for
the results of the present analysis and for their
combination with those previously published [4], 
resulting in 95\% 
C.L. lower limits on each model parameter. 
Both in the case of the excited electron and of the string mass scale,
the values given for $\rm \xi$ correspond to setting 
$\lambda_\gamma$ and $|\lambda|$ to 1.\label{tab_lim}}
\end{table}}


\end{document}